\documentclass[pre,twocolumn,showpacs]{revtex4}
\usepackage{epsfig}
\usepackage{amssymb}

\begin{document}

\title{Numerical study of Fermi-Pasta-Ulam recurrence for water waves over finite depth}
\author{V. P. Ruban}
\email{ruban@itp.ac.ru}
\affiliation{Landau Institute for Theoretical Physics,
2 Kosygin Street, 119334 Moscow, Russia} 

\date{\today}

\begin{abstract}
Highly accurate direct numerical simulations have been performed for two-dimensional 
free-surface potential flows of an ideal incompressible fluid over a constant depth 
$h$, in the gravity field $g$. In each numerical experiment, at $t=0$ the free 
surface profile was in the form $y=A_0\cos(2\pi x/L)$, and the velocity field 
${\bf v}=0$. 
The computations demonstrate the phenomenon of Fermi-Pasta-Ulam (FPU) recurrence
takes place in such systems for moderate initial wave amplitudes 
$A_0\lesssim 0.12 h$ and spatial periods at least $L\lesssim 120 h$. 
The time of recurrence $T_{\rm FPU}$ is well fitted by the formula
$T_{\rm FPU}(g/h)^{1/2}\approx 0.16(L/h)^2(h/A_0)^{1/2}$.
\end{abstract}


\maketitle

Many nonlinear dispersive waves are known to exhibit the Fermi-Pasta-Ulam (FPU)
recurrence, when a (finite-size) system nearly repeats its initial state after 
some period of evolution. For the first time, this phenomenon was observed in 
the famous numerical experiment \cite{FPU1955} with one-dimensional (1D) lattices of
nonlinear oscillators. The impact of that 
observation on the subsequent development of nonlinear science was very deep. 
In particular, it resulted in the discovery of solitons by Zabusky and Kruskal
\cite{ZK1965}.  Since then, the FPU recurrence and related phenomena were studied 
in many physical contexts (see, e.g., 
\cite{Z1973,T1977,YF1978,Infeld1981,YL1982,AK1986,OOSB1998,SEH2001,ZLY2003,Zabusky2005,BI2005}, 
and references therein). 

In general, the FPU recurrence takes place in a system if its dynamics is nearly integrable. 
Two different reasons can result in such approximate integrability.
In the first case, just a few (2-3) collective degrees of freedom are effectively involved into the evolution due to a small system size and low level of nonlinearity, and that restricted
dynamics is integrable (for example, such variant of FPU recurrence takes place for 
3D deep-water waves \cite{YL1982,ZLY2003}). 
In the second possible case, the evolution is governed by equations of motion which 
are close to some completely integrable system \cite{Z1973}. 
The first case does not require any additional symmetry and therefore is more common, 
while the second case is more interesting, since it allows for a large number of 
degrees of freedom to be involved into the process.

As the theory of water waves is concerned, there are three the most popular integrable 
models for three essentially different dynamic regimes: 
(i) the nonlinear Schroedinger equation (NLSE) approximates an envelope of 
deep-water waves, 
(ii) the Korteweg-de-Vries (KdV) equation describes weakly nonlinear dispersive
unidirectional shallow-water waves, 
and (iii) the Boussinesq equations approximately describe bidirectional 
shallow-water waves (concerning integrability of the Boussinesq equations, see \cite{Kaup1975,Kupershmidt1985,Smirnov1986,BZ2002,ZhangLi2003}). 
While the FPU recurrence was investigated in detail for the regimes (i)
\cite{T1977,YF1978,Infeld1981,YL1982,AK1986,ZLY2003} and (ii) \cite{OOSB1998}, 
it was not studied for the regime (iii), 
which case includes water waves propagating in a closed flume. 
The present work is intended to fill this gap by means of direct numerical simulations.

\begin{figure}
\begin{center}
   \epsfig{file=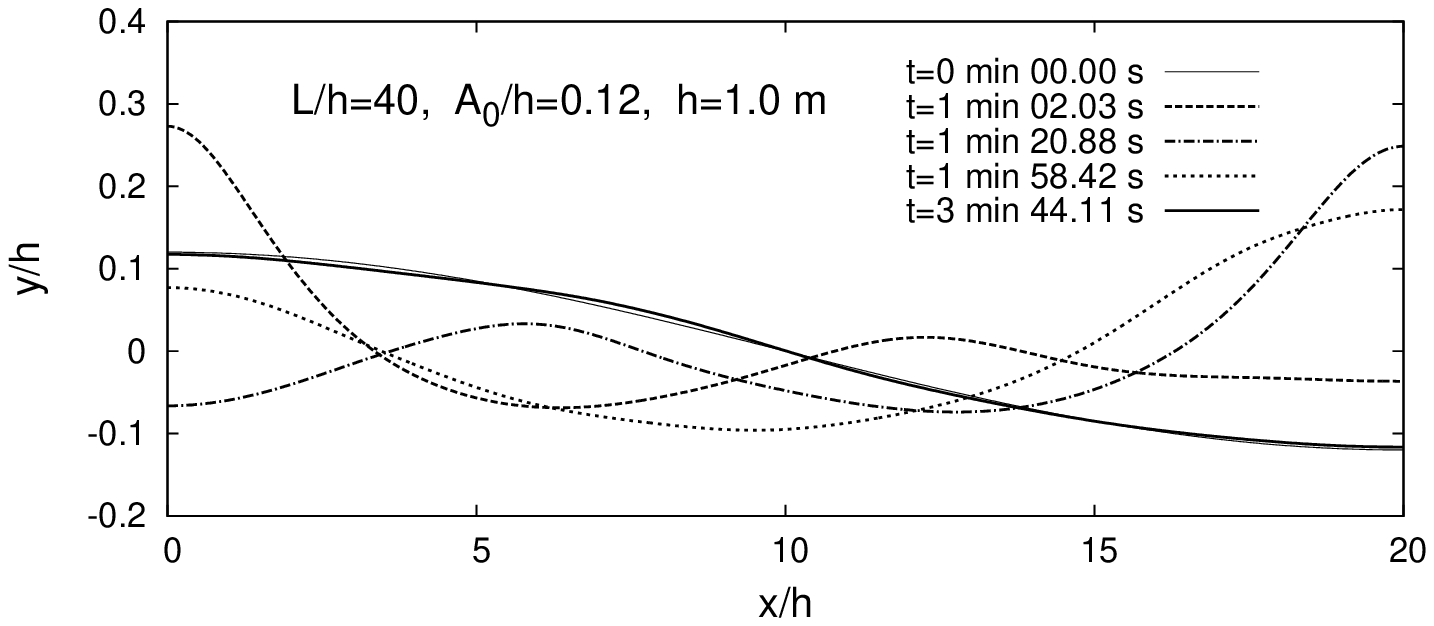,width=80mm}\\
   \epsfig{file=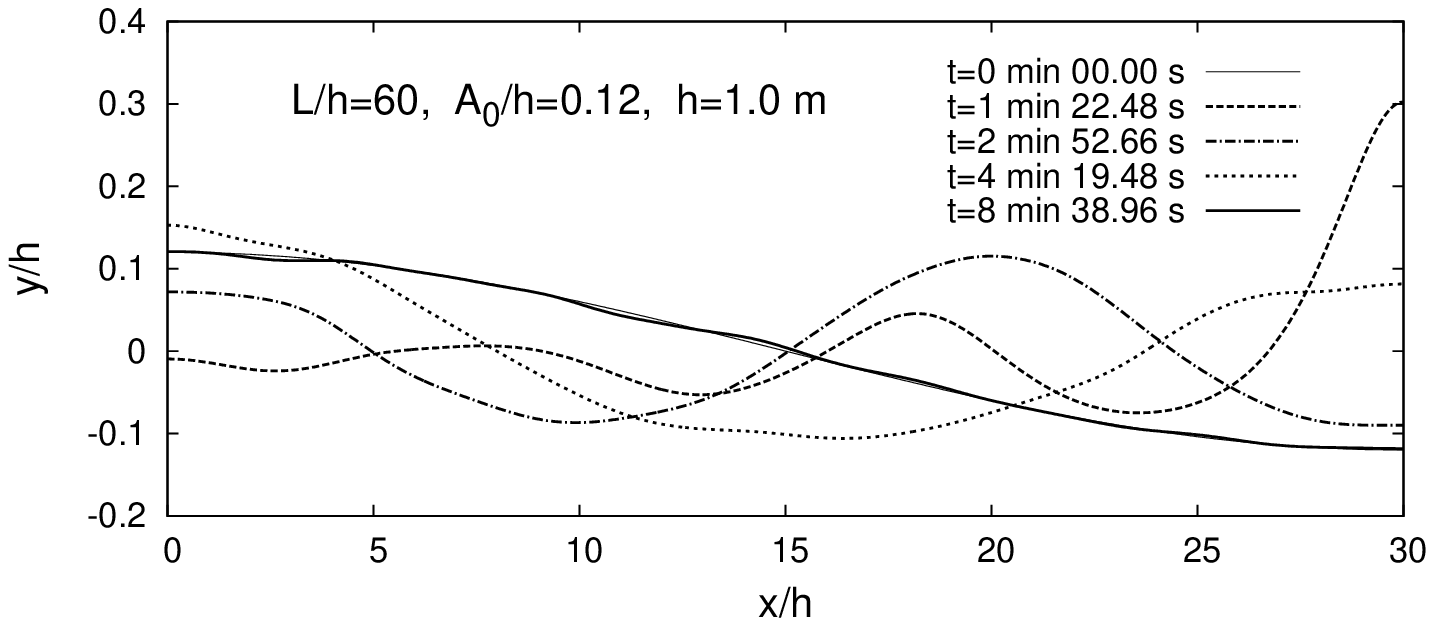,width=80mm}\\
   \epsfig{file=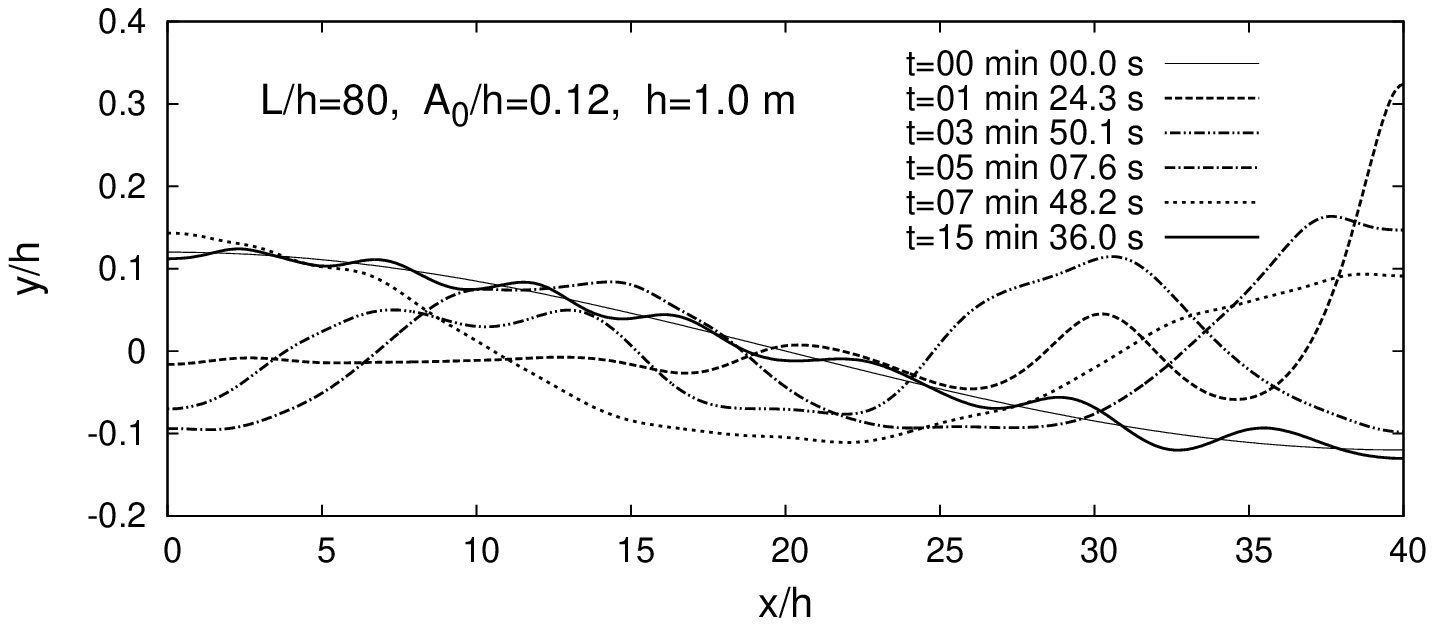,width=80mm}
\end{center}
\caption{Three examples of FPU recurrence in water wave systems: 
the free surface profiles at $t=0$ and at $t=T_{\rm FPU}(h=1\, {\rm m};L;A_0=0.12h)$ 
are very close for $L/h=40$ and for $L/h=60$; for $L/h=80$ the recurrence
is not so perfect but still apparent. The curves with the highest elevation
correspond to the moments when a ``soliton'' collides with the wall.} 
\label{FPU_recurrence} 
\end{figure}

\begin{figure}
\begin{center}
   \epsfig{file=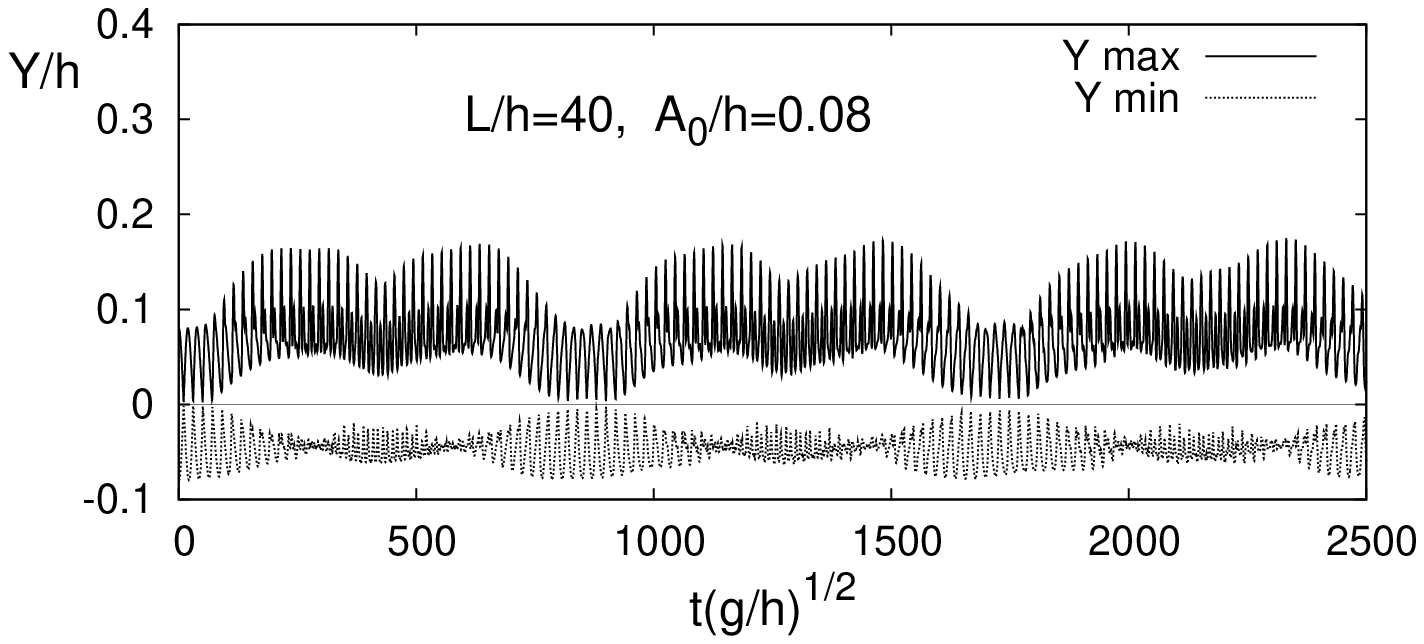,width=85mm}\\
   \epsfig{file=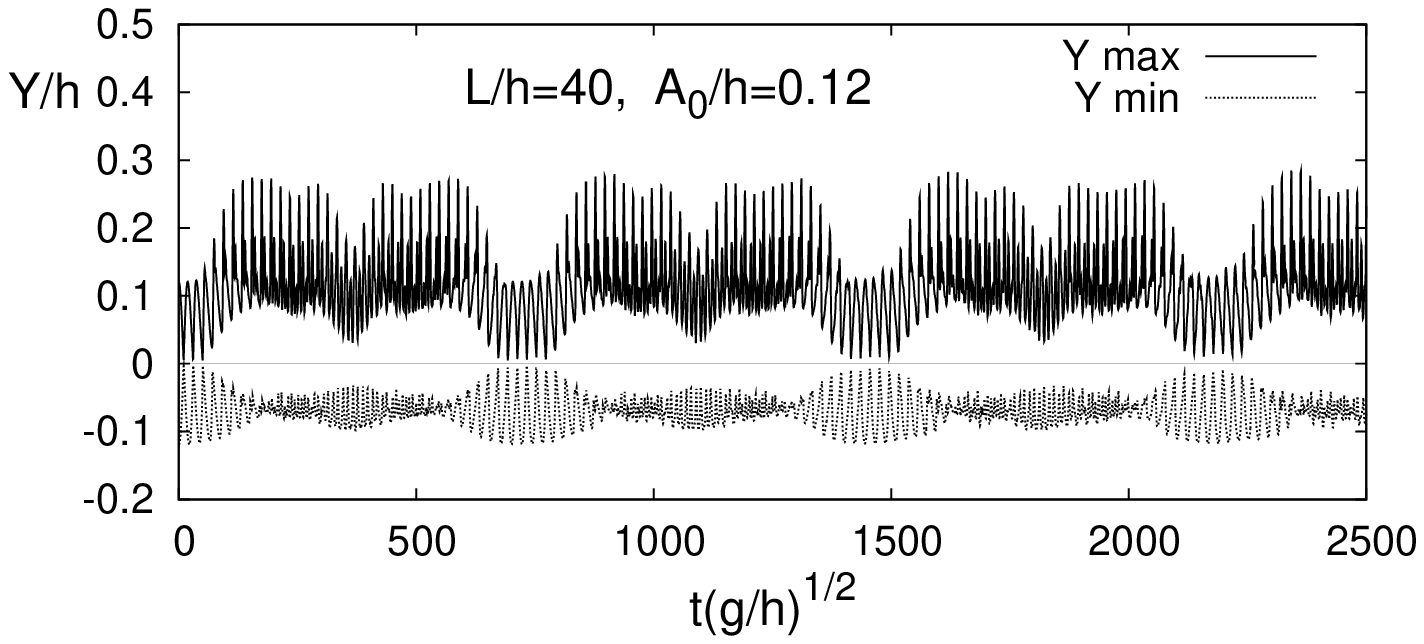,width=85mm}\\
   \epsfig{file=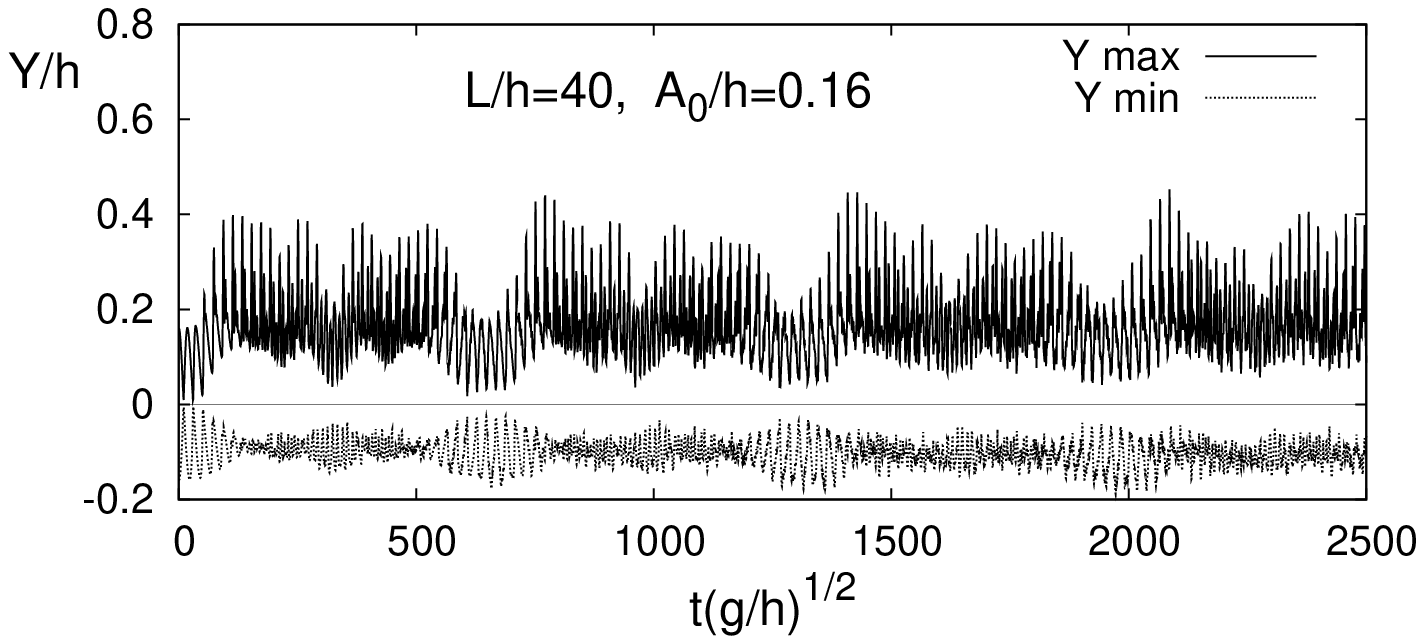,width=85mm}\\
   \epsfig{file=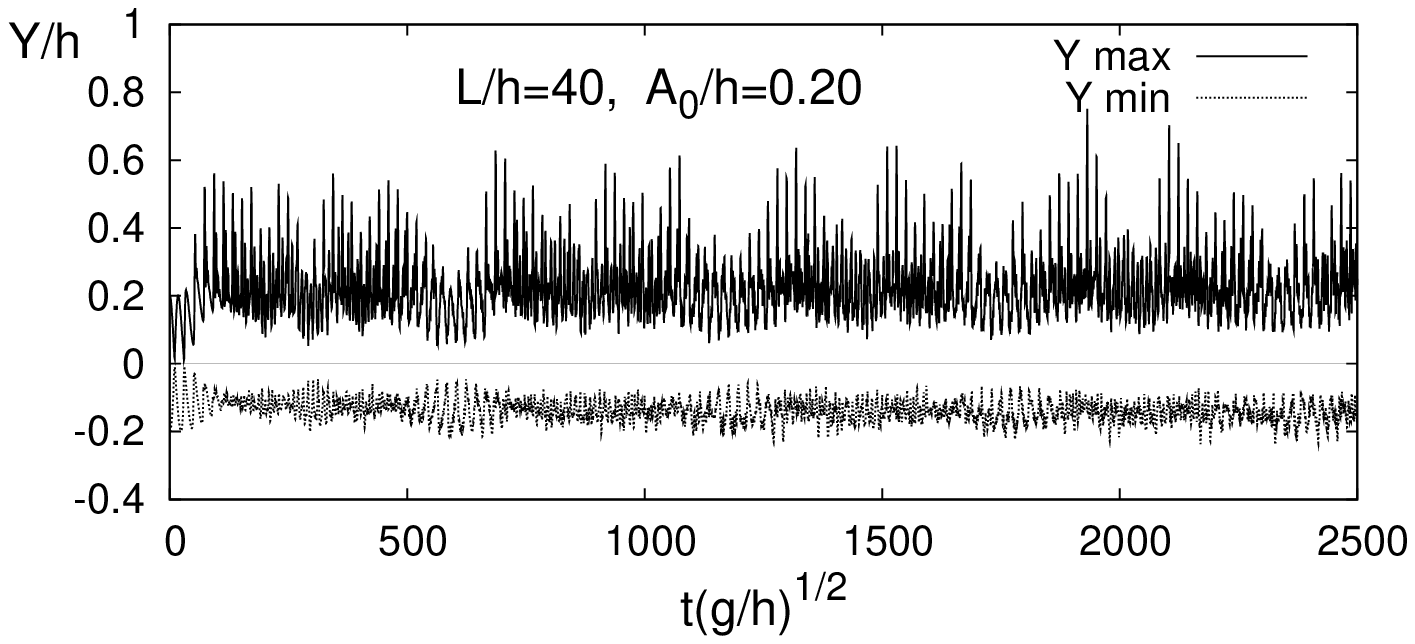,width=85mm}\\
   \epsfig{file=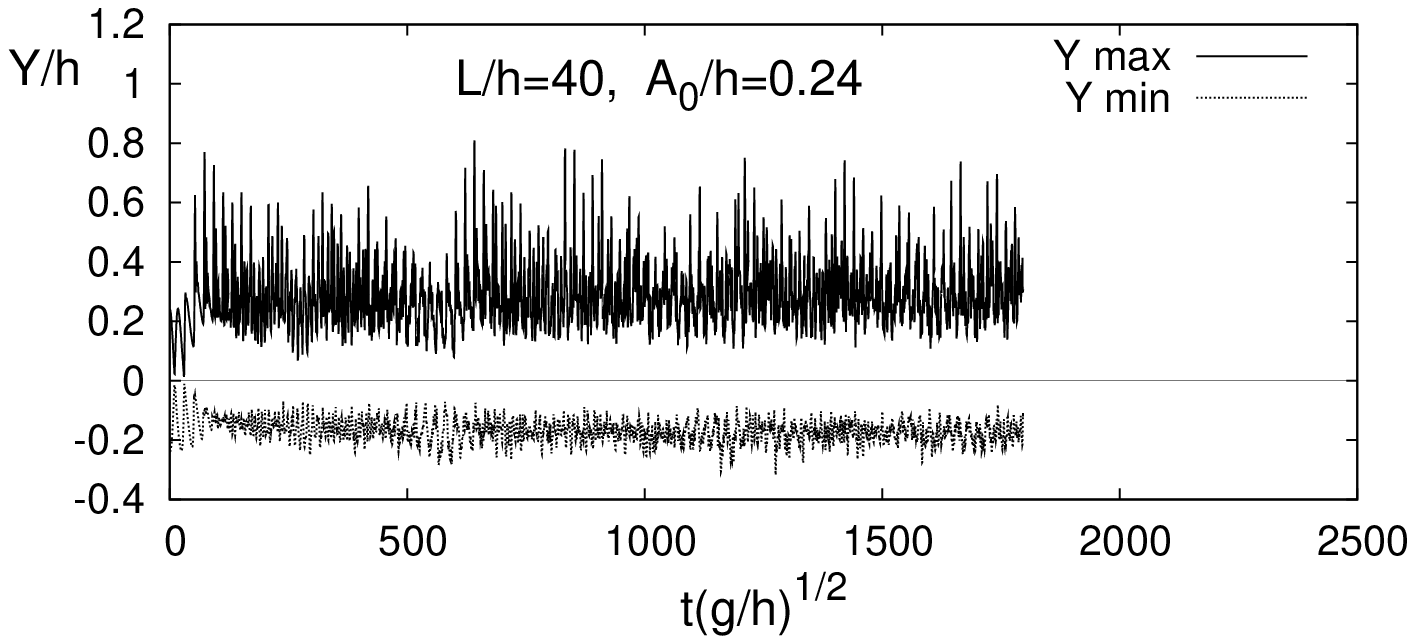,width=85mm} 
\end{center}
\caption{The maximum and minimum elevations of the free boundary 
versus time for $L/h=40$ and different initial amplitudes. The peaks of the
dependences at approximately regular time intervals 
correspond to collisions of ``solitons'' with the ``walls''.} 
\label{L40} 
\end{figure}

\begin{figure}
\begin{center}
   \epsfig{file=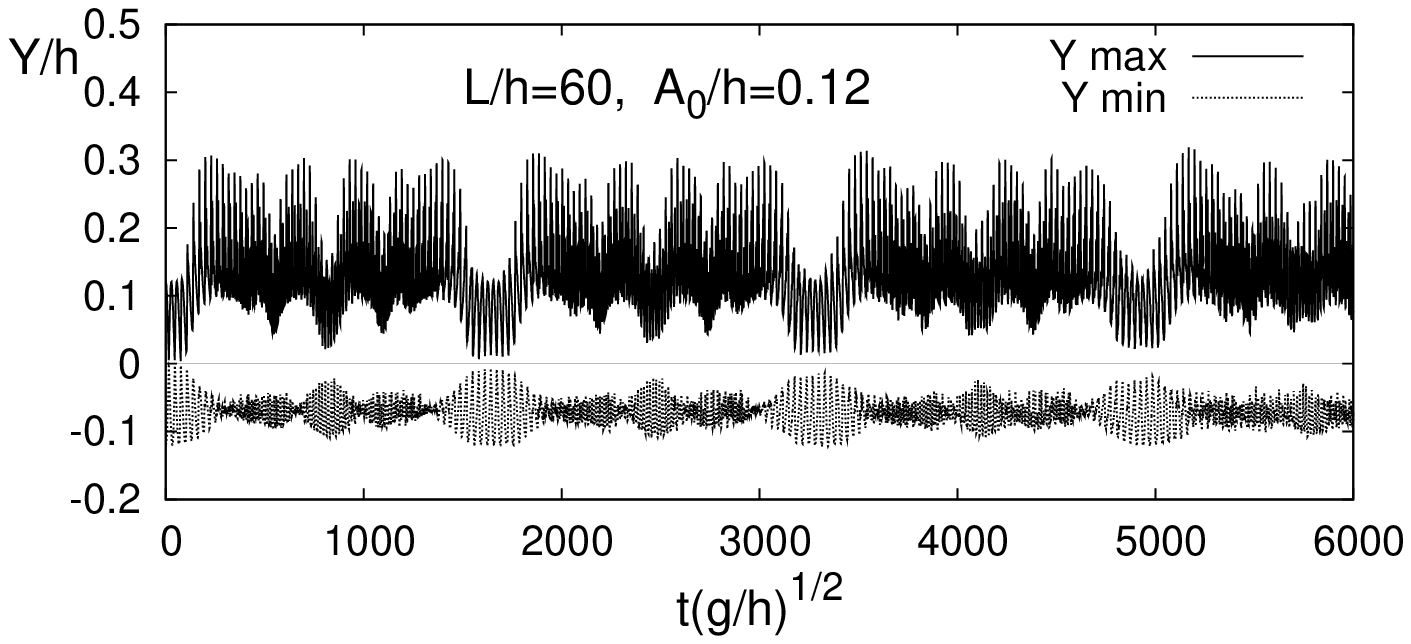,width=85mm}\\
   \epsfig{file=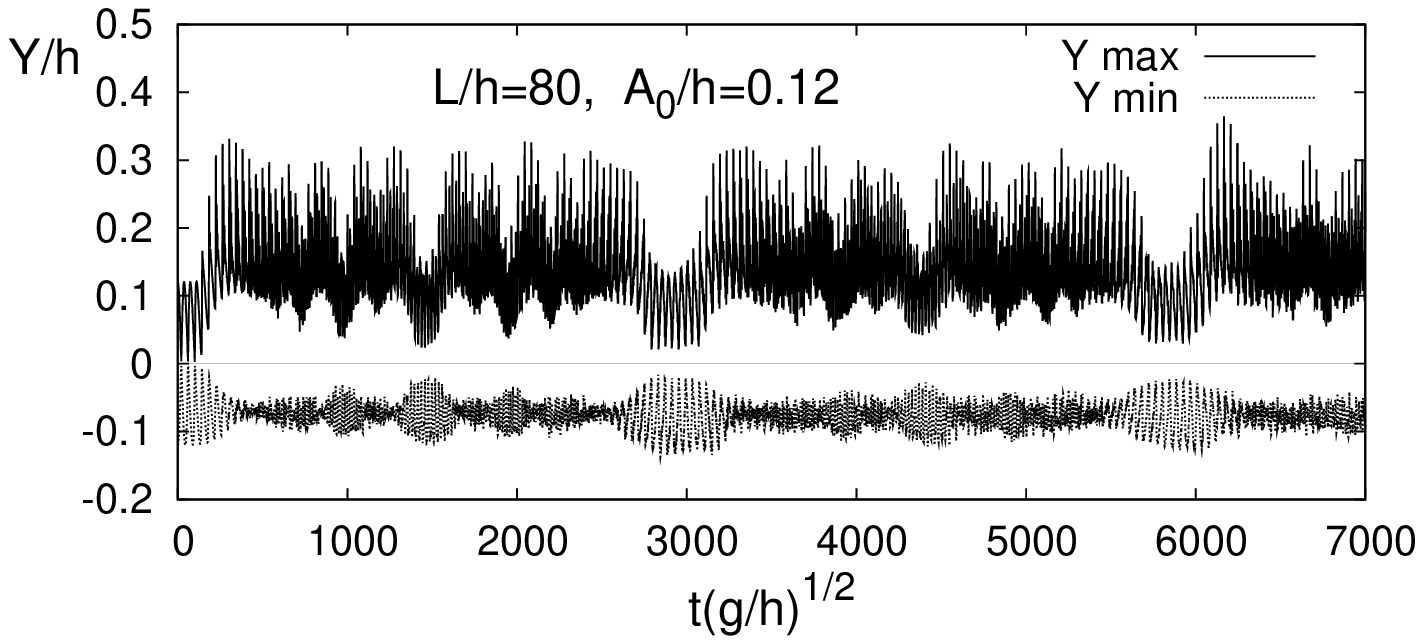,width=85mm}\\
   \epsfig{file=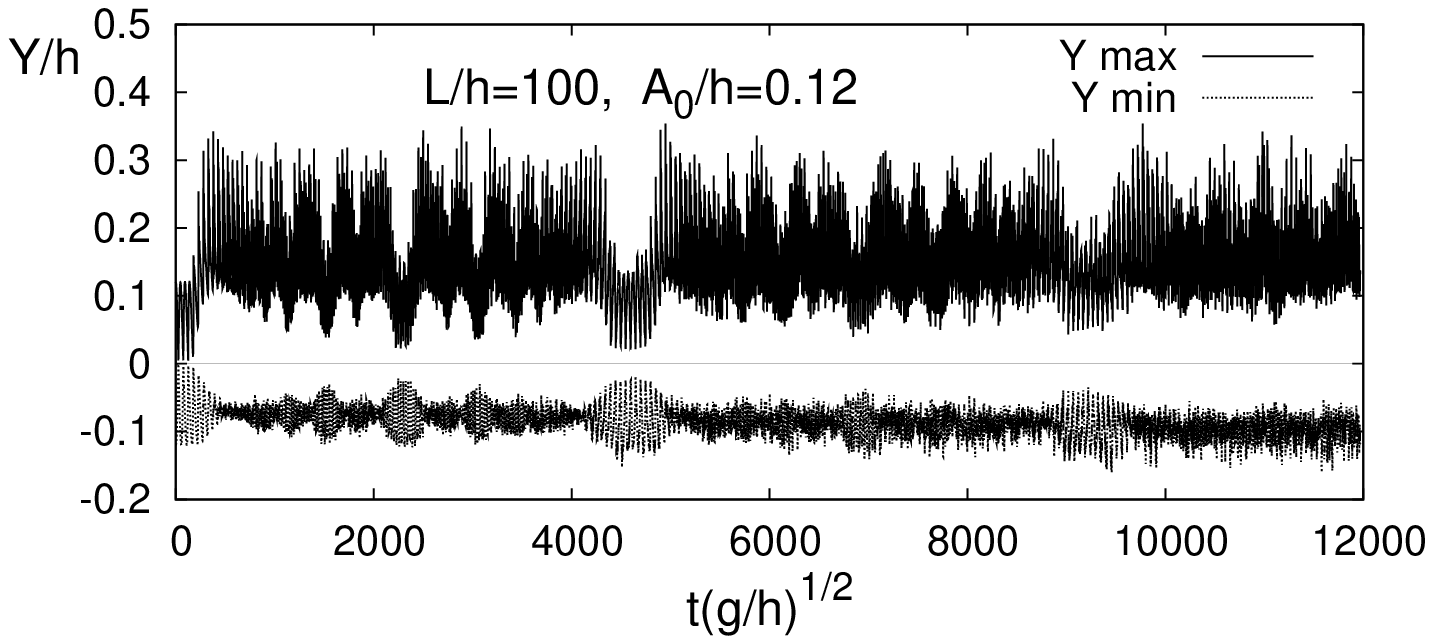,width=85mm}
\end{center}
\caption{FPU recurrence for $A_0/h=0.12$ and different $L$.
With increasing $L$, the dynamics between the recurrence cases becomes
more and more complicated: additional minima in the envelope of $Y_{max}$ appear,
which correspond to significant excitation of the second, or the third, or a 
higher Fourier mode in nearly standing-wave regime.} 
\label{A012} 
\end{figure}

It is important that nonlinear dynamics of free water surface is not exactly
integrable (see, e.g., \cite{OOSB1998,CGHHS2006}, and references therein), and therefore 
it was not obvious \emph{a priori} in what parameter region the FPU recurrence 
can be observed. To clarify this question, here the highly accurate
numerical method based on exact equations of motion for two-dimensional 
(2D) free-surface potential flows of an ideal incompressible fluid 
was employed \cite{R2004PRE,R2008PRE}, in the most 
simple variant when the bottom is horizontal at a constant depth $h$. 
In each numerical experiment, at $t=0$ the free surface profile was taken 
in the form $y=A_0\cos(2\pi x/L)$ (the gravity acceleration $g$ 
is directed against $y$ axis), and the velocity field was zero everywhere. 
Thus, this configuration corresponds to waves  in a basin
with vertical walls at $x=0$ and $x=L/2$.

Since the exact equations of motion are written in terms of so called conformal
variables, and the surface shape is given in a parametric form 
$X+iY=-ih+(1+i\hat R)\rho(\vartheta,t)$, 
with $\hat R$ being a linear integral operator \cite{R2004PRE, R2008PRE},
it was a nontrivial task how to determine the real function $\rho(\vartheta,0)$
corresponding to a given initial surface profile $y=\eta_0(x)$. This technical
problem was solved by an auxiliary numerical procedure, when in one of the two
evolutionary equations (namely, in the Bernoulli equation) the combination 
$\{\psi_t+g Y\}$ was temporarily replaced with the combination 
$\{\psi_t+g[Y-\eta_0(x)] + \Gamma\psi\}$, where $\psi(\vartheta,t)$ is the surface
value of the velocity potential, and $\Gamma=const>0$ is some artificial damping. 
With this modification, the free surface quickly evolved from $Y=0$ to the required 
initial profile, and after that we turned to the original equations 
\cite{R2004PRE, R2008PRE}.

The parameters $A_0$ and $L$ were taken inside the  region $0.04\le A_0/h\le 0.24$
and $20\le L/h\le 120$. The computations have demonstrated that
the phenomenon of FPU recurrence indeed takes place 
for moderate initial wave amplitudes $A_0\lesssim 0.12 h$.
After a few oscillations in a standing-wave regime, the system enters a regime
with soliton-like coherent structures moving between the ``vertical walls''.
The shapes of the ``solitons'' evolve with time, and after a period $T_{\rm FPU}$, 
which depends on $h$, $L$, and $A_0$, the system again enters the standing-wave 
regime and approximately repeats the initial state, as it is shown in Fig.1.
With a realistic value $h\sim1$ m, the period of recurrence is several minutes. 

\begin{figure}
\begin{center}
   \epsfig{file=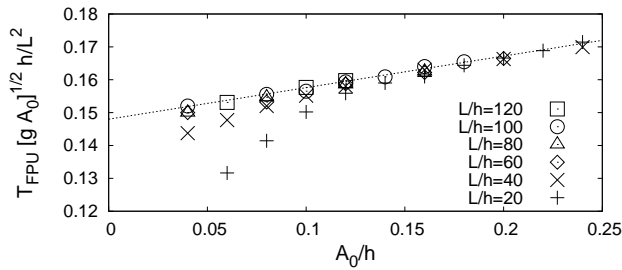,width=85mm}\\
\end{center}
\caption{The time of recurrence, excluding the cases when $(A_0/h)^{1/2}(L/h)\lesssim 1$, 
is well fitted by the formula
$T_{\rm FPU}(g/h)^{1/2}\approx C(L/h)^2(h/A_0)^{1/2}$, with $C=0.148+0.096(A_0/h)$.
Roughly, $C= 0.16\pm 0.01$.} 
\label{T_FPU} 
\end{figure}

In Fig.2 and Fig.3, the maximum and minimum elevations of the free boundary 
versus time are shown for different $A_0$ and $L$. These dependences 
clearly indicate the presence of the FPU phenomenon. 
However, a ``quality'' of the recurrence is not uniform in the parametric region: 
it is better for smaller initial amplitudes and shorter spatial periods $L$.
For large amplitudes $A_0/h\gtrsim 0.20$, one can hardly recognize more 
than one recurrence case. In that strongly nonlinear regime, very high waves 
grow in the system, which can eventually produce sharp crests and break (not shown). 

The time of recurrence rapidly increases with the spatial period $L$, 
and behavior of the system between the recurrences becomes more complicated, 
since a larger number of solitons participates in the dynamics, 
$N_{\rm s}\sim(L/h)(A_0/h)^{1/2}$. 
In the course of evolution, there are some time intervals when a significant 
part of the energy is concentrated in the second, or in the third, 
or in a higher Fourier mode.

The numerical results for $T_{\rm FPU}$ for different $L$ and $A_0$ 
are summarized in Fig.4, which shows that, excluding the cases when 
$(L/h)(A_0/h)^{1/2}\lesssim 1$ (under this condition, the length $L$ 
becomes too small to contain a soliton with an amplitude about $A_0$), 
the time of recurrence is well fitted by a formula
\begin{equation}
T_{\rm FPU}\left(\frac{g}{h}\right)^{1/2}\approx \left[0.148+0.096\frac{A_0}{h}\right]
\left(\frac{L}{h}\right)^2\left(\frac{h}{A_0}\right)^{1/2}.
\end{equation}
It should be noted here that in each case the quantity $T_{\rm FPU}$ was determined
rather roughly, just by looking at graphs like those shown in Fig.2 and Fig.3.
Therefore the corresponding accuracy is rather low, 
despite the fact that the accuracy of the simulations is very high
(so, the energy conservation was at least 7-8 decimal digits from the beginning 
of each computation to its end). 

In order to observe the FPU recurrence in a real-world flume filled with water, 
the viscous friction near the bottom and near the side walls should be taken into account. 
Thus an additional condition arises, $\gamma\cdot T_{\rm FPU}\ll 1$, 
where the damping coefficient $\gamma$ can be estimated as
\begin{equation}
\gamma\lesssim\frac{1}{h}\sqrt{\nu\sqrt{\frac{g}{h}}}\sim h^{-5/4},
\end{equation}
where $\nu$ is the kinematic viscosity.
Since with fixed ratios $A_0/h$ and $L/h$ the time of recurrence behaves as $h^{1/2}$,
the product $\gamma \cdot T_{\rm FPU}$ behaves as $h^{-3/4}$, and therefore it 
can be indeed small with sufficiently large $h$.

To conclude, in the present work the phenomenon of Fermi-Pasta-Ulam recurrence
was observed in the numerical experiments modeling potential one-dimensional 
water waves in a closed basin. Based on the numerical results, 
a simple fitting formula for the period of recurrence is suggested.

These investigations were supported by RFBR grant 09-01-00631,
by the ``Leading Scientific Schools of Russia'' grant 6885.2010.2,
and by the Program ``Fundamental Problems of Nonlinear Dynamics'' 
from the RAS Presidium.

\end{document}